\documentclass[
  aps,
  prl,
  reprint,
  superscriptaddress,
  nofootinbib
]{revtex4-2}

\usepackage{amsmath,amssymb}
\usepackage{bm}
\usepackage{graphicx}
\usepackage{hyperref}
\usepackage{xcolor}
\usepackage{float}
 \usepackage{ulem} 
\colorlet{BLUE}{blue}

\newcommand{\bx}{\bm{x}}
\newcommand{\by}{\bm{y}}

\newcommand{\bnu}{\bm{\nu}}
\newcommand{\bF}{\bm{F}}

\begin{document}

\title{Information-Theoretic Characterization of Macroscopic Chaos Emerging from the Chemical Master Equation}

\author{Kenshin Matsumoto}
\author{Shin-ichi Sasa}
\affiliation{Department of Physics, Kyoto University, Kyoto 606-8502, Japan}

\date{\today}

\begin{abstract}
Open chemical reaction networks exhibit stochastic concentration dynamics at finite system sizes, whereas their macroscopic limit is governed by deterministic rate equations that can display chaos. In this Letter, we show theoretically that a rate of information loss constructed from two-time mutual information recovers the Kolmogorov-Sinai entropy in the deterministic limit. We verify this result through numerical simulations of a Markov jump process for a three-species system involving seven reactions.
\end{abstract}

\maketitle

% Main text begins here.

{\it Introduction.---}
Open chemical reaction networks, which exchange matter with particle reservoirs, can exhibit irregular concentration dynamics on macroscopic scales.  
In the standard macroscopic description, these dynamics are governed by deterministic rate equations that can exhibit chaos, as demonstrated in both theoretical and experimental studies of chemical reaction systems \cite{TomitaTsuda1979,GyorgyiField1992,PetrovEtAl1993}.
At the molecular level, however, reactions occur as discrete random events.  A finite-volume reaction network is therefore fundamentally described by a chemical master equation, or equivalently by a Markov jump process for the numbers of chemical species \cite{vanKampen,Gillespie}. 
This stochastic description remains applicable even when the corresponding macroscopic rate equation exhibits chaos and has been used to investigate intrinsic fluctuations in such systems \cite{GeysermansNicolis1993,WangLi1998}.
The standard deterministic notion of chaos does not apply directly to such a process.  Relating macroscopic chemical chaos to this stochastic description is thus analogous in spirit to deriving thermodynamic behavior from molecular statistical mechanics.

Two established facts suggest how this gap can be bridged.
First, for density-dependent reaction networks, the molecular population vector rescaled by system size converges in the large-system-size limit to a trajectory governed by the deterministic rate equation~\cite{Kurtz1970,Kurtz1972}.
%At the next order, fluctuations around this trajectory are described by a Gaussian approximation obtained from the system-size expansion~\cite{Kurtz1978,vanKampen}.
Second, the information generated by a deterministic chaotic trajectory is quantified by the Kolmogorov--Sinai (KS) entropy. 
More precisely, a partition of phase space assigns a symbolic sequence to each trajectory, and the KS entropy is the supremum of the Shannon information rate of this sequence over all such partitions \cite{EckmannRuelle1985}.  
These facts lead to a concrete question: is there an information-theoretic quantity defined entirely in terms of the Markov jump process that converges to the KS entropy of the macroscopic rate equation?

Connections among dynamical instability, stochasticity, and information production have been studied from several perspectives.  
Finite-resolution entropy rates were developed to distinguish deterministic chaos from noise \cite{GaspardWang}.
Rates of entropy change were related to the KS entropy for chaotic maps~\cite{LatoraBaranger} and to the sum of the positive generalized Lyapunov exponents for environment-coupled chaotic Hamiltonian systems~\cite{Pattanayak}.
Random-perturbation approaches sought to recover deterministic dynamical entropy from weakly stochastic
dynamics \cite{Ostruszka}. 
Pesin-type entropy formulas have also been established for classes of smooth random dynamical systems~\cite{LedrappierYoung1988}.
For chemical systems, a stochastic approach to entropy production in chemical chaos was developed in Ref.~\cite{Gaspard2020}.  
These works demonstrate that information-theoretic quantities remain meaningful for characterizing dynamical instability beyond the direct analysis of deterministic trajectories.
To the best of our knowledge, however, they do not identify a quantity defined directly from a chemical master equation whose large-system-size limit yields the KS entropy of the corresponding rate equation.

In this Letter, we show that the rate at which two-time mutual information decreases provides such a quantity. 
Let $\bm{X}_t^{(V)}$ denote the stochastic concentration vector at time $t$ for a chemical system of size $V$, with $\bm{X}_0^{(V)}$ drawn from its stationary distribution.
For the mutual information $I(\bm{X}_t^{(V)};\bm{X}_0^{(V)})$,
we establish that the asymptotic rate of decrease $-\mathrm{d}I(\bm{X}_t^{(V)};\bm{X}_0^{(V)})/\mathrm{d}t$ converges to the KS entropy $h_{\mathrm{KS}}$ of the deterministic rate equation.

The mechanism behind this result is transparent in the system-size expansion.  
Stationarity allows the loss of mutual information to be expressed exactly as the growth of the Shannon entropy of $\bm{X}_t^{(V)}$ conditioned on $\bm{X}_0^{(V)}$. 
Before a crossover time $t_\mathrm{c}(V)$ that diverges as $V\to \infty$, the conditional fluctuations are locally Gaussian, and their covariance evolves under the linearized deterministic flow.  
The fluctuation widths grow exponentially along unstable directions while remaining bounded along stable directions at leading exponential order.  
Consequently, the conditional-entropy growth rate is the sum of the positive Lyapunov exponents.
Pesin's equality then identifies this sum with $h_{\mathrm{KS}}$ \cite{EckmannRuelle1985,Pesin}.  

\textit{Setup.---}
To investigate the relation between information loss in stochastic dynamics and deterministic chaos, we consider a chemical reaction network consisting of $N$ chemical species $X_1,\ldots,X_N$ and $R$ reaction channels indexed by $r=1,\ldots,R$.
For channel $r$, let $\alpha_{ir}$ and $\beta_{ir}$ denote the stoichiometric coefficients of species $X_i$ on the reactant and product sides, respectively, and define the stoichiometric vector $\bnu_r\in\mathbb{Z}^N$ by $(\bnu_r)_i=\beta_{ir}-\alpha_{ir}$.

As a concrete example throughout this Letter, we consider a modified version of the CS14 chemical reaction system proposed by Plesa and Sprott~\cite{PlesaSprott2026}.
This model has $N=3$ chemical species and $R=7$ reaction channels:
\begin{align}
  r=1:\quad Y&\longrightarrow X+Y,
  &
  r=2:\quad 2X&\longrightarrow3X,
  \notag\\
  r=3:\quad X+Z&\longrightarrow2Z,
  &
  r=4:\quad Z&\longrightarrow Y,
  \notag\\
  r=5:\quad X+Y&\longrightarrow X,
  &
  r=6:\quad 3X&\longrightarrow2X,
  \notag\\
  r=7:\quad \varnothing&\longrightarrow Z.
  \label{eq:modified-cs14}
\end{align}
The added reactions $r=6$ and $r=7$ suppress large excursions of $X$ and prevent absorption at finite $V$, respectively.

Let $V$ denote the system size and $\bm{n}$ the vector of molecular populations.
We write $\bx=\bm{n}/V$ for a possible concentration value and denote the stochastic concentration vector at time $t$ by $\bm{X}_t^{(V)}$.
Each occurrence of reaction channel $r$ changes the concentration as $\bx\mapsto\bx+\bnu_r/V$ at transition rate $Va_r^{(V)}(\bx)$.
Under mass-action kinetics, the finite-$V$ rate function is
\begin{equation}
  a_r^{(V)}(\bx)
  =
  k_r
  \prod_{i=1}^{N}
  V^{-\alpha_{ir}}
  \left[
    Vx_i(Vx_i-1)\cdots
    (Vx_i-\alpha_{ir}+1)
  \right].
  \label{eq:finite-volume-rate}
\end{equation}
As $V\to\infty$, it converges to
\begin{equation}
  a_r(\bx)
  =
  \lim_{V\to\infty}a_r^{(V)}(\bx)
  =
  k_r\prod_{i=1}^{N}x_i^{\alpha_{ir}}.
  \label{eq:macroscopic-rate}
\end{equation}
Let $p_t^{(V)}(\bx)$ denote the probability distribution of the stochastic concentration at time $t$. Its time evolution is governed by
\begin{align}
  \frac{d}{dt}p_t^{(V)}(\bx)
  =
  V\sum_{r=1}^{R}
  \Biggl[
    &
    a_r^{(V)}
    \!\left(\bx-\frac{\bnu_r}{V}\right)
    p_t^{(V)}
    \!\left(\bx-\frac{\bnu_r}{V}\right)
    \notag\\[-0.5ex]
    &{}-
    a_r^{(V)}(\bx)
    p_t^{(V)}(\bx)
  \Biggr].
  \label{eq:master}
\end{align}
We assume $\lim_{\|\bx\|\to\infty}p_t^{(V)}(\bx)=0$.

\begin{figure}[t]
  \centering
  \includegraphics[
    width=\linewidth
  ]{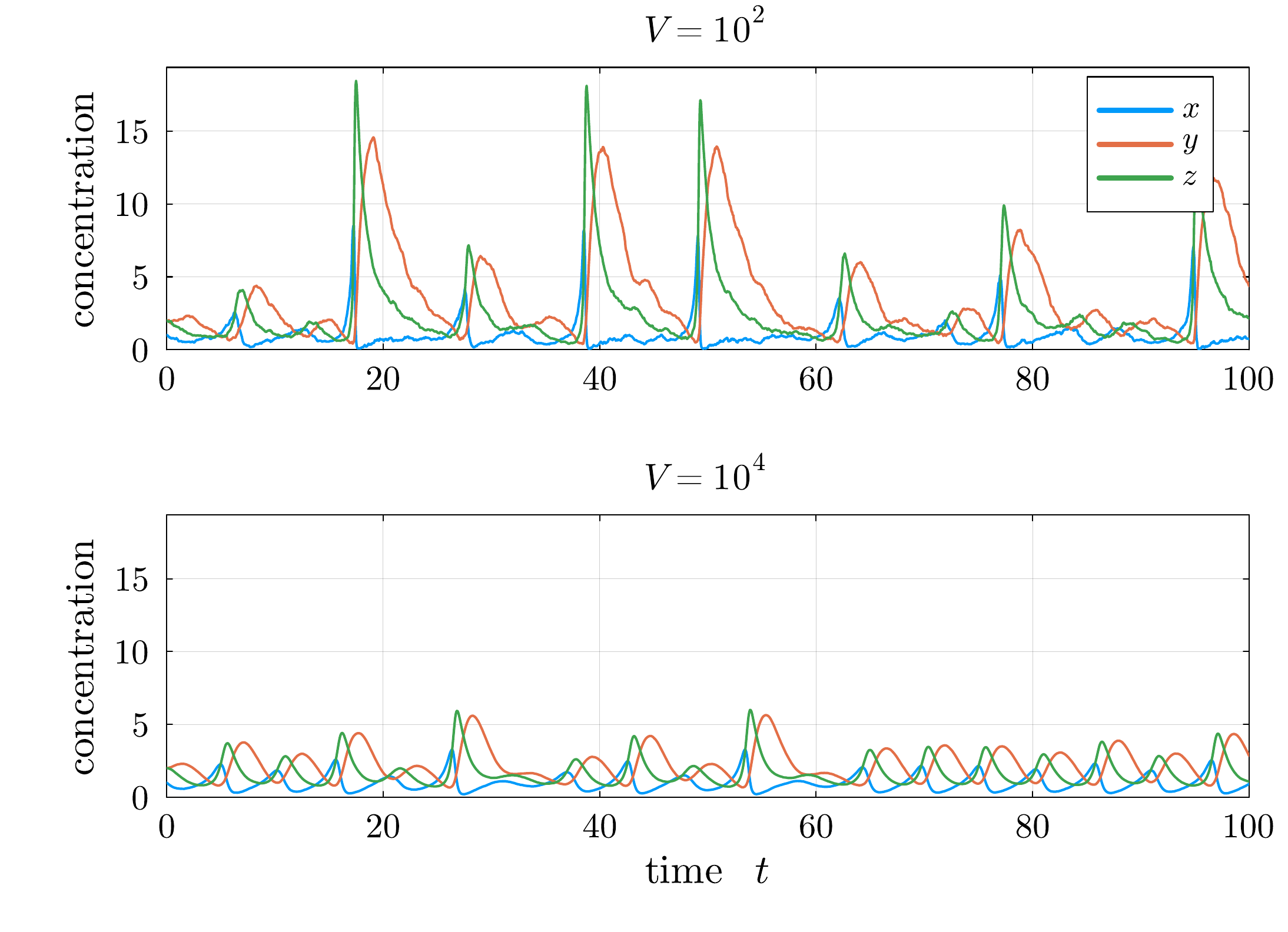}
  \caption{
    Representative concentration time series of the modified CS14 model generated using the tau-leap method.
    The concentrations $(x(t),y(t),z(t))$ are shown for $V=10^2$ (top) and $V=10^4$ (bottom).
  }
  \label{fig:tau-leap-timeseries}
\end{figure}

For the modified CS14 model, $\bx=(x,y,z)^{\mathsf T}$ and the finite-$V$ rate functions are
\begin{equation}
  \begin{aligned}
    a_1^{(V)}(\bx)
    &=
    k_1y,
    &
    a_2^{(V)}(\bx)
    &=
    k_2x\left(x-\frac{1}{V}\right),
    \\
    a_3^{(V)}(\bx)
    &=
    k_3xz,
    &
    a_4^{(V)}(\bx)
    &=
    k_4z,
    \\
    a_5^{(V)}(\bx)
    &=
    k_5xy,
    &
    a_6^{(V)}(\bx)
    &=
    k_6 x
    \left(x-\frac{1}{V}\right)
    \left(x-\frac{2}{V}\right),
    \\
    a_7^{(V)}(\bx)
    &=
    k_7.
  \end{aligned}
  \label{eq:modified-cs14-rates}
\end{equation}
Together with the stoichiometric vectors in Eq.~\eqref{eq:modified-cs14}, these functions define the master equation of the modified CS14 model through Eq.~\eqref{eq:master}.
The rate constants for channels $r=1,\ldots,7$ are $k_1,\ldots,k_7$, respectively, with $k_1=0.2$, $k_2=k_3=k_4=k_5=1$, $k_6=0.02$, and $k_7=0.01$.
Figure~\ref{fig:tau-leap-timeseries} compares representative stochastic trajectories $\hat{\bx}_t$ generated by the tau-leap method~\cite{Gillespie2001}.
Reaction fluctuations produce large irregular excursions at $V=10^2$, whereas they are reduced and the concentration trajectories become smoother at $V=10^4$, illustrating the approach to the deterministic limit.

In the deterministic limit $V\to\infty$, the stochastic trajectory $\hat{\bx}_t$ converges on finite time intervals to the deterministic trajectory $\bx(t)$ satisfying
\begin{equation}
  \frac{d\bx}{dt}
  =
  \sum_{r=1}^{R}
  \bnu_r a_r(\bx)
  \equiv
  \bF(\bx),
  \label{eq:rate-equation}
\end{equation}
where $a_r(\bx)$ is defined in Eq.~\eqref{eq:macroscopic-rate}.
For the modified CS14 model, this rate equation becomes
\begin{align}
  \frac{dx}{dt}
  &=
  k_1y+k_2x^2-k_3xz-k_6 x^3,
  \notag\\
  \frac{dy}{dt}
  &=
  k_4z-k_5xy,
  \notag\\
  \frac{dz}{dt}
  &=
  k_3xz-k_4z+k_7.
  \label{eq:modified-cs14-rate-equation}
\end{align}
For the parameters above, Eq.~\eqref{eq:modified-cs14-rate-equation} exhibits chaotic dynamics, whose orbital instability we characterize by its Lyapunov spectrum.

Let $\bm{\phi}_t(\bx_0)$ denote the solution of Eq.~\eqref{eq:rate-equation} starting from $\bx_0$, and let $\bm{M}_t(\bx_0)$ satisfy
$\dot{\bm{M}}_t=D\bm{F}(\bm{\phi}_t(\bx_0))\bm{M}_t$
with $\bm{M}_0=\bm{I}$, where
$[D\bm{F}(\by)]_{ij}\equiv\partial F_i(\by)/\partial y_j$.
Let $\sigma_1(t)\geq\cdots\geq\sigma_N(t)$
be the singular values of $\bm{M}_t(\bx_0)$.
The Lyapunov spectrum is defined by
\begin{equation}
\lambda_i
=
\lim_{t\to\infty}
\frac{1}{t}
\ln\sigma_i(t),
\qquad
i=1,\ldots,N.
\label{eq:lyapunov-spectrum}
\end{equation}
The KS entropy $h_{\mathrm{KS}}$ is defined as the supremum, over all partitions of phase space, of the growth rate of the Shannon entropy of the induced symbolic sequences. Although the appropriate partition is not known a priori, Pesin's identity states that, under suitable conditions, $h_{\mathrm{KS}}$ equals the sum of the positive Lyapunov exponents. Numerical integration of the rate equation and its variational equation gives the Lyapunov spectrum $(\lambda_1,\lambda_2,\lambda_3)=(0.065,0,-0.86)$. The positive maximal exponent confirms chaotic dynamics and, through Pesin's identity, gives $h_{\mathrm{KS}}=0.065$.

The question we address is whether an information-theoretic quantity defined for the stochastic dynamics governed by Eq.~\eqref{eq:master} converges to the KS entropy in the macroscopic limit $V\to\infty$.

{\it Key Quantity.---}
An important observation is that the KS entropy quantifies the rate at which a symbolic sequence recorded over a time interval $[0,t]$ gains information about the initial state at $t=0$. Equivalently, it quantifies the rate at which a measurement of the state at time $t$ loses information about the initial state at $t=0$. The latter information is quantified by the mutual information $I(\bm{X}_t^{(V)};\bm{X}_0^{(V)})$:
\begin{align}
  &I\!\left(
    \bm{X}_t^{(V)};
    \bm{X}_0^{(V)}
  \right)
  \equiv
  \sum_{\bx_0,\bx}
  p_{0,t}^{(V)}(\bx_0,\bx)
  \ln
  \frac{
    p_{0,t}^{(V)}(\bx_0,\bx)
  }{
    p_0^{(V)}(\bx_0)
    p_t^{(V)}(\bx)
  },
  \label{eq:mutual-information}
\end{align}
where $p_{0,t}^{(V)}(\bx_0,\bx)$ denotes their joint probability distribution. Mutual information can be defined for any stochastic system and is nonincreasing in time for any Markov process because the data-processing inequality gives
\begin{equation}
    I\!\left(\bm{X}_t^{(V)};\bm{X}_0^{(V)}\right)
  \leq
  I\!\left(\bm{X}_s^{(V)};\bm{X}_0^{(V)}\right),
  \label{eq:data-processing}
\end{equation}
for any $0\leq s\leq t$~\cite{CoverThomas2006}. 
A derivation is provided in the End Matter.
The heuristic identification of the KS entropy with the rate of information lossmotivates the following conjecture in the macroscopic limit $V\to\infty$:
\begin{align}
-\frac{d}{dt}I(\bm{X}_t^{(V)};\bm{X}_0^{(V)}) = h_{\rm KS}.
\label{main}
\end{align}

\begin{figure}[t]
  \centering
  \includegraphics[
    width=\linewidth
  ]{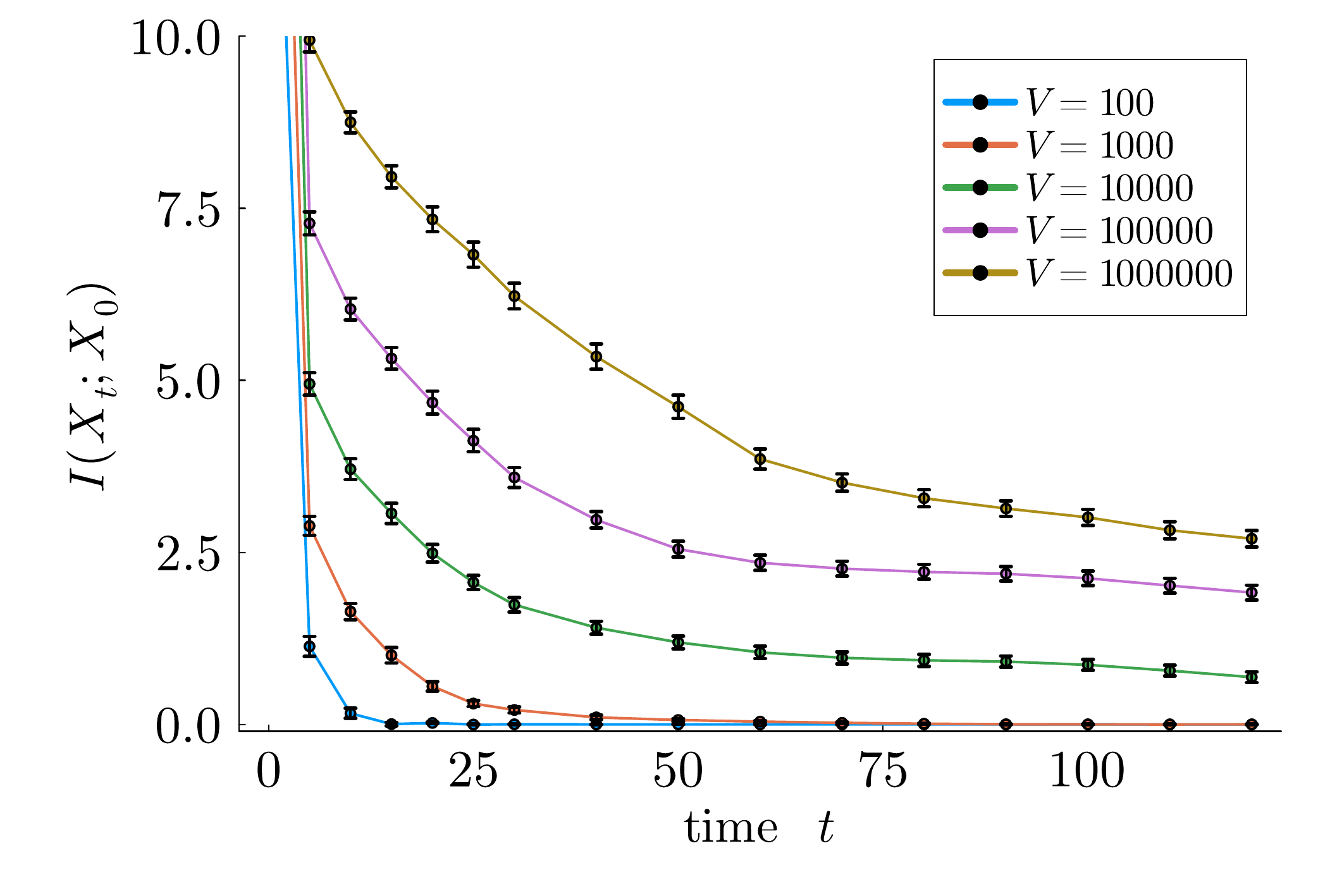}
  \caption{
    Time evolution of $I(\bm{X}_t^{(V)};\bm{X}_0^{(V)})$ for the modified CS14 model at $V=10^j$, $j=2,\ldots,6$.
    Symbols show $k$-nearest-neighbor estimates with $k=10$, and error bars indicate standard errors.
    The number of initial concentrations is $N_{\mathrm{ini}}=50$, and the number of sample trajectories generated from each initial concentration is $N_{\mathrm{sample}}=10^6$.
  }
  \label{fig:mutual-information-system-size}
\end{figure}

For the modified CS14 model, we sample the initial concentration from the stationary distribution and compute $I(\bm{X}_t^{(V)};\bm{X}_0^{(V)})$ for $V=10^j$ with $j=2,\ldots,6$.
Sample trajectories are generated by the tau-leap method~\cite{Gillespie2001}, and the mutual information is estimated from the pairs $(\bm{X}_0^{(V)},\bm{X}_t^{(V)})$ using the $k$-nearest-neighbor estimator with $k=10$~\cite{KraskovEtAl2004}.
Figure~\ref{fig:mutual-information-system-size} shows the resulting time dependence.
At small system sizes, the mutual information decreases rapidly as reaction fluctuations become stronger.
As $V$ increases, an intermediate-time regime of approximately linear decrease emerges and extends progressively to later times.

We next show that the rate of information loss in this linear regime converges to the KS entropy in the macroscopic limit and clarify the conditions under which the conjecture in Eq.~\eqref{main} holds.

{\it Main Result.---}
We rewrite the mutual information as
\begin{equation}
  I\!\left(
    \bm{X}_t^{(V)};
    \bm{X}_0^{(V)}
  \right)
  =
  H\!\left(
    \bm{X}_t^{(V)}
  \right)
  -
  H\!\left(
    \bm{X}_t^{(V)}
    \mid
    \bm{X}_0^{(V)}
  \right).
  \label{eq:mutual-information-entropy}
\end{equation}
Because the initial concentration is sampled from the stationary distribution, the marginal distribution of $\bm{X}_t^{(V)}$ and hence its Shannon entropy $H(\bm{X}_t^{(V)})$ are independent of time $t$.
The time dependence of the mutual information is therefore determined entirely by the conditional entropy
$H(\bm{X}_t^{(V)}\mid\bm{X}_0^{(V)})$, which has the exact representation
\begin{equation}
H\!\left(\bm{X}_t^{(V)}\mid\bm{X}_0^{(V)}\right)
=
\sum_{\bx_0}\!p_{\mathrm{ss}}^{(V)}(\bx_0)
\!H\!\left(\bm{X}_t^{(V)}\mid\bm{X}_0^{(V)}=\bx_0\right),
\label{eq:conditional-entropy-transition}
\end{equation}
where $H\!(\bm{X}_t^{(V)}\mid\bm{X}_0^{(V)}=\bx_0)$ denotes the Shannon entropy of the conditional distribution $p_t^{(V)}(\bx\mid\bx_0)$ of the concentration $\bx$ at time $t$, given the initial concentration $\bx_0$.

We estimate $H\!(\bm{X}_t^{(V)}\mid\bm{X}_0^{(V)}=\bx_0)$ for a fixed $\bx_0$. 
At early times and for large $V$, $p_t^{(V)}(\bx\mid\bx_0)$ is well approximated by a Gaussian distribution centered on the deterministic trajectory $\bm{\phi}_t(\bx_0)$ and having covariance matrix $V^{-1}\bm{\Sigma}_t(\bx_0)$.
This Gaussian description breaks down around the crossover timescale $t_{\mathrm c}(V)$ because the fluctuation amplitude grows as $V^{-1/2}e^{\lambda_1t}$, where $t_{\mathrm c}(V)$ is estimated from
\begin{equation}
V^{-1/2}e^{\lambda_1t_{\mathrm c}(V)}
= \mathcal{O}(V^0).
\label{eq:crossover-time}
\end{equation}
Within this Gaussian approximation, the conditional entropy is given by the Shannon entropy of the Gaussian distribution, denoted by $H_{\mathrm G}(\bm{X}_t^{(V)}\mid\bm{X}_0^{(V)}=\bx_0)$. We then obtain
\begin{align}
&H_{\mathrm G}\!\left(
\bm{X}_t^{(V)}
\mid
\bm{X}_0^{(V)}=\bx_0
\right)
= C +
\frac{1}{2}
\ln\det\bm{\Sigma}_t(\bx_0),
\label{eq:gaussian-conditional-entropy-fixed}
\end{align}
where $C$ is independent of $t$ but depends on $N$ and $V$.

To remove the $V$ dependence, we introduce a fixed reference time $t_{\mathrm{ref}}>0$ 
and define the Gaussian conditional-entropy increment by
\begin{align}
\Delta H_{\mathrm G}^{(V)}(t;t_{\mathrm{ref}})
&\equiv
H_{\mathrm G}\!\left(
\bm{X}_t^{(V)}
\mid
\bm{X}_0^{(V)}
\right)
-
H_{\mathrm G}\!\left(
\bm{X}_{t_{\mathrm{ref}}}^{(V)}
\mid
\bm{X}_0^{(V)}
\right).
\label{eq:gaussian-entropy-increment-definition}
\end{align}
The time-independent terms in Eq.~\eqref{eq:gaussian-conditional-entropy-fixed} cancel between the two terms on the right-hand side of Eq.~\eqref{eq:gaussian-entropy-increment-definition}, yielding
\begin{equation}
\Delta H_{\mathrm G}^{(V)}(t;t_{\mathrm{ref}})
=
\frac{1}{2}
\sum_{\bx_0}
p_{\mathrm{ss}}^{(V)}(\bx_0)
\ln
\frac{
\det\bm{\Sigma}_t(\bx_0)
}{
\det\bm{\Sigma}_{t_{\mathrm{ref}}}(\bx_0)
}.
\label{eq:finiteV-gaussian-information-loss}
\end{equation}
The long-time behavior of $\Delta H_{\mathrm G}^{(V)}(t;t_{\mathrm{ref}})$ is derived explicitly in the End Matter. In essence, decomposing the covariance evolution into the linearized deterministic flow and accumulated reaction fluctuations shows that the latter cancel the contributions from the negative Lyapunov exponents.
The remaining growth rate is therefore $\sum_{\lambda_i(\bx_0)>0}\lambda_i(\bx_0)$.
We thus obtain
\begin{equation}
\lim_{t\to\infty}
\frac{
\Delta H_{\mathrm G}^{(V)}(t;t_{\mathrm{ref}})
}{
t-t_{\mathrm{ref}}
}
=
\sum_{\bx_0}p_{\mathrm{ss}}^{(V)}(\bx_0)
\sum_{\lambda_i(\bx_0)>0}
\lambda_i(\bx_0).
\label{eq:covariance-lyapunov-rate}
\end{equation}

Recalling Eq.~\eqref{eq:mutual-information-entropy}, we define
the relative information loss by
\begin{equation}
\Delta I^{(V)}(t;t_{\mathrm{ref}})
\equiv
I\!\left(
\bm{X}_{t_{\mathrm{ref}}}^{(V)};
\bm{X}_0^{(V)}
\right)
-
I\!\left(
\bm{X}_t^{(V)};
\bm{X}_0^{(V)}
\right).
\label{eq:relative-information-loss}
\end{equation}
For $0<t_{\mathrm{ref}}<t\ll t_{\mathrm c}(V)$, 
the Gaussian approximation to the conditional distribution gives
\begin{equation}
\Delta I^{(V)}(t;t_{\mathrm{ref}})
=
\Delta H_{\mathrm G}^{(V)}(t;t_{\mathrm{ref}}).
\label{eq:finiteV-gaussian-validity}
\end{equation}
Because the crossover time $t_{\mathrm c}(V)$ diverges as the system size increases,
Eq.~\eqref{eq:finiteV-gaussian-validity} relates the large-system-size limit of the relative information loss to the Gaussian entropy increment. Here, we assume that the finite-$V$ stationary probability measures converge to an invariant measure $\mu$ of the deterministic dynamics.
Taking $V\to\infty$ at fixed $t$ in Eq.~\eqref{eq:finiteV-gaussian-validity}, using the convergence of $p_{\mathrm{ss}}^{(V)}$ to $\mu$ in Eq.~\eqref{eq:covariance-lyapunov-rate}, and subsequently applying Pesin's identity in the long-time limit, we obtain
\begin{equation}
\lim_{t\to\infty}
\lim_{V\to\infty}
\frac{
\Delta I^{(V)}(t;t_{\mathrm{ref}})
}{
t-t_{\mathrm{ref}}
}
=
h_{\mathrm{KS}}.
\label{eq:universal-information-loss}
\end{equation}
If $\mathrm{d}I(\bm{X}_t^{(V)};\bm{X}_0^{(V)})/\mathrm{d}t$ has a well-defined limit as $t\to\infty$ after $V\to\infty$, Eq.~\eqref{main} follows in this order of limits.

\begin{figure}[t]
  \centering
  \includegraphics[
    width=\linewidth
  ]{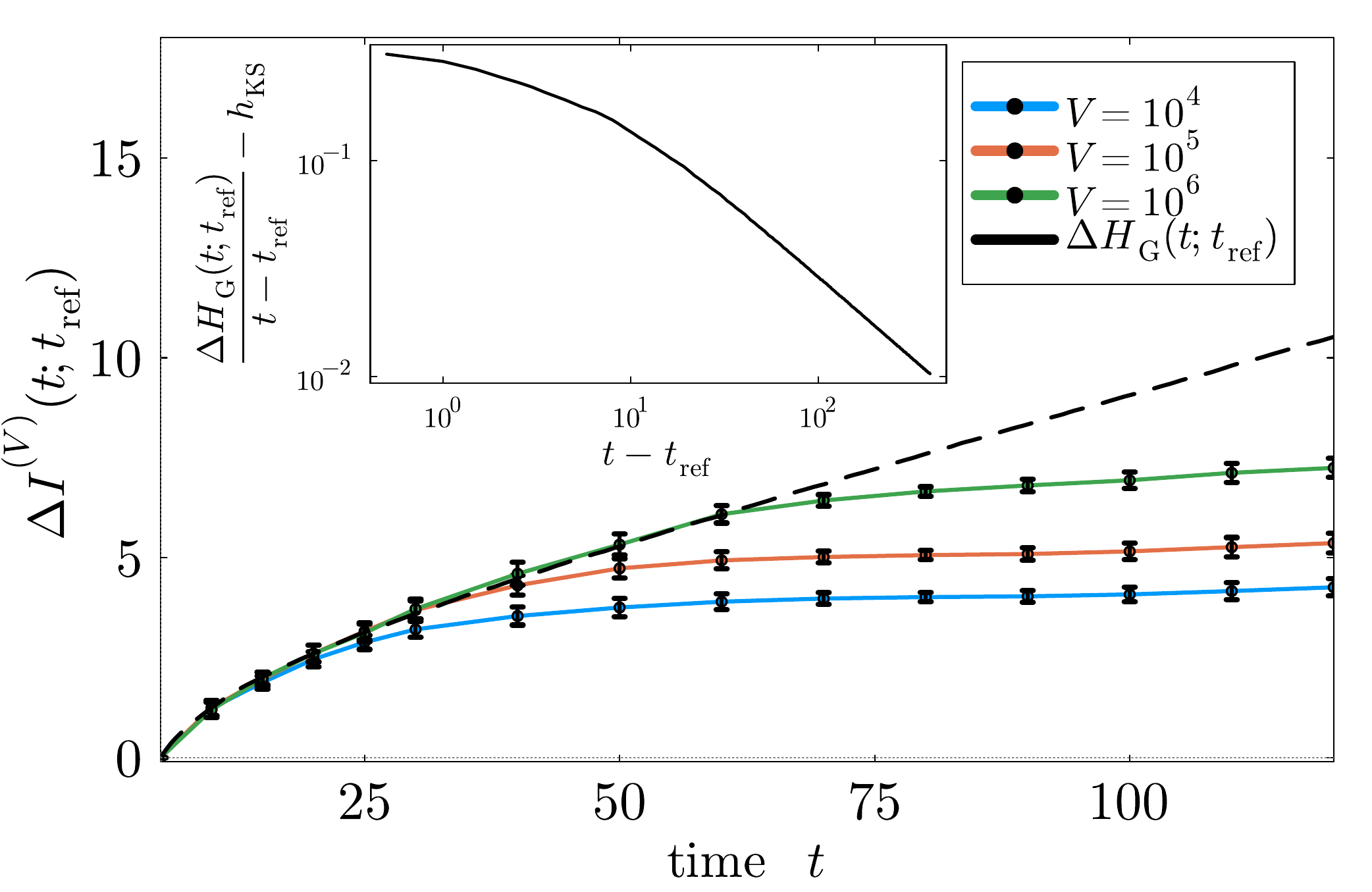}
  \caption{
    Relative information loss $\Delta I^{(V)}(t;t_{\mathrm{ref}})$ from $t_{\mathrm{ref}}=5$ for $V=10^4,10^5$, and $10^6$.
    Symbols show kNN estimates obtained using $10^6$ samples for each system size with $k=10$.
    The error bars indicate standard errors, and the colored lines connect the symbols.
    The black dashed curve shows the Gaussian entropy increment in Eq.~\eqref{eq:finiteV-gaussian-information-loss}, obtained by integrating the covariance equation~\eqref{eq:linear-covariance}.
    The inset shows the residual $\Delta H_{\mathrm G}^{(V)}(t;t_{\mathrm{ref}})/(t-t_{\mathrm{ref}})-h_{\mathrm{KS}}$ on logarithmic axes.
  }
  \label{fig:relative-information-loss}
\end{figure}
We confirm these theoretical predictions numerically for the modified CS14 model.
We evaluate the relative information loss $\Delta I^{(V)}(t;t_{\mathrm{ref}})$ for $V=10^4,10^5$, and $10^6$ using $t_{\mathrm{ref}}=5$.
Figure~\ref{fig:relative-information-loss} demonstrates that the finite-$V$ information loss $\Delta I^{(V)}(t;t_{\mathrm{ref}})$ initially follows the Gaussian entropy increment $\Delta H_{\mathrm G}^{(V)}(t;t_{\mathrm{ref}})$ in Eq.~\eqref{eq:finiteV-gaussian-information-loss} for all three system sizes, as predicted by Eq.~\eqref{eq:finiteV-gaussian-validity}.
Each numerical curve eventually departs from the Gaussian prediction, and the interval of agreement extends to later times as $V$ increases.
The inset of Fig.~\ref{fig:relative-information-loss} shows that the finite-time rate of Gaussian information loss $\Delta H_{\mathrm G}^{(V)}(t;t_{\mathrm{ref}})/(t-t_{\mathrm{ref}})$ approaches $h_{\mathrm{KS}}$ at long times, in accordance with Eq.~\eqref{eq:covariance-lyapunov-rate}.
These numerical results demonstrate that the Gaussian approximation describes the information loss over an increasingly long time interval as the system size increases, while its asymptotic slope approaches the KS entropy, as predicted by Eq.~\eqref{eq:universal-information-loss}.

{\it Concluding remarks.---}
We formulated the loss rate of two-time mutual information for chemical master equations and showed that this quantity recovers the KS entropy in the deterministic limit. Before concluding this Letter, we offer a few remarks.

Although we focus on chemical master equations in this Letter, our formulation can be applied to general stochastic systems that admit a deterministic limit. One example is a chaotic system subject to weak noise. In such a system, the loss rate of two-time mutual information can likewise be shown to recover the KS entropy in the deterministic limit and may therefore provide a new tool for analyzing chaotic systems. Furthermore, the loss rate of two-time mutual information can be considered for any stochastic system, including probabilistic cellular automata and lattice-gas models, in which the discrete nature persists even in the deterministic limit. It would be interesting to characterize chaotic behavior by measuring this rate.

An extension of our work to quantum chaos may offer another direction. According to current understanding, the exponential growth of an operator commutator, as measured by an out-of-time-ordered correlator (OTOC), typically probes the maximal Lyapunov exponent~\cite{MaldacenaEtAl2016,RozenbaumEtAl2017} in the semiclassical limit. 
Goldfriend and Kurchan defined a quantum KS entropy as the entropy production rate induced by coupling the system to a weak auxiliary bath and derived a quantum Pesin relation connecting this rate to phase-space expansion~\cite{GoldfriendKurchan2021}.
Cao formulated the statistics of weak measurements of macroscopic quantum fluctuations in terms of multi-time correlation and response functions and related them to the dynamical entropy of the measurement ancillas~\cite{Cao2026}.
These studies introduce an auxiliary bath or measurement apparatus and characterize the resulting entropy production or measurement statistics.
The present work takes a complementary approach based on the two-time distribution of an autonomous classical Markov process, whose stochasticity arises from finite-size reaction events.
The relation between the present classical result and these quantum constructions remains to be clarified.

{\it Acknowledgements.---}
We thank J. Kurchan for helpful discussions and for bringing Refs.~\cite{GoldfriendKurchan2021,Cao2026} to our attention.
This work was supported by JSPS KAKENHI (Grant Nos.~JP25K00923 and JP26H00383) and JST SPRING (Grant No.~JPMJSP2110).

{\it Data Availability.---}
The data that support the findings of
this article are openly available~\cite{CS14InformationLossCode}.

\clearpage
\onecolumngrid
\begin{center}
  {\large\bfseries End Matter\par}
\end{center}
\twocolumngrid

\setcounter{equation}{0}
\renewcommand{\theequation}{E\arabic{equation}}
\renewcommand{\theHequation}{E\arabic{equation}}
%\textit{Derivation of Eq.~\eqref{eq:data-processing}.---}
\section{Derivation of \NoCaseChange{Eq.~\eqref{eq:data-processing}}}

We first review the data-processing inequality,
\begin{equation}
I(X;Z)\leq I(X;Y),
\label{eq:endmatter-generic-data-processing}
\end{equation}
for random variables $X$, $Y$, and $Z$ forming the Markov chain~\cite{CoverThomas2006}
\begin{equation}
X\longrightarrow Y\longrightarrow Z.
\label{eq:endmatter-generic-markov-chain}
\end{equation}
Let $p(x,y,z)$ be the joint probability distribution of $X$, $Y$, and $Z$.
Using
\begin{equation}
\frac{p(x,y,z)}{p(x)p(y,z)}=\frac{p(z|x,y)p(x|y)p(y)}{p(x)p(z|y)p(y)},
\end{equation}
we obtain
\begin{equation}
I(X;Y,Z)
=
I(X;Y)
+
I(X;Z\mid Y).
\label{eq:endmatter-chain-rule-first}
\end{equation}
Exchanging $Y$ and $Z$ gives
\begin{equation}
I(X;Y,Z)
=
I(X;Z)
+
I(X;Y\mid Z).
\label{eq:endmatter-chain-rule-second}
\end{equation}
Combining Eqs.~\eqref{eq:endmatter-chain-rule-first} 
and \eqref{eq:endmatter-chain-rule-second} and using 
the Markov property $I(X;Z\mid Y)=0$, 
we obtain
\begin{equation}
I(X;Y)
=
I(X;Z)
+
I(X;Y\mid Z).
\label{eq:endmatter-data-processing-identity}
\end{equation}
Because conditional mutual information is nonnegative,
Eq.~\eqref{eq:endmatter-data-processing-identity} yields
Eq.~\eqref{eq:endmatter-generic-data-processing}.

For the stochastic process considered here, we set
\begin{equation}
X=\bm{X}_0^{(V)},
\qquad
Y=\bm{X}_s^{(V)},
\qquad
Z=\bm{X}_t^{(V)}
\label{eq:endmatter-variable-correspondence}
\end{equation}
for any $0\leq s\leq t$. The data-processing inequality in
Eq.~\eqref{eq:endmatter-generic-data-processing} then yields
Eq.~\eqref{eq:data-processing} in the main text.

We further note that the initial and current concentrations become statistically
independent in the long-time limit at fixed $V$ if the Markov process is mixing and
has a unique stationary distribution. This implies
\begin{equation}
\lim_{t\to\infty}
I\!\left(
\bm{X}_t^{(V)};
\bm{X}_0^{(V)}
\right)
=0.
\label{eq:endmatter-finiteV-mixing}
\end{equation}

%\textit{Derivation of Eq.~\eqref{eq:covariance-lyapunov-rate}.---}
\section{Derivation of \NoCaseChange{Eq.~\eqref{eq:covariance-lyapunov-rate}}}

In this section, we derive 
\begin{equation}
\lim_{t\to\infty}
\frac{1}{2t}
\ln\det\bm{\Sigma}_t(\bx_0)
=
\sum_{\lambda_i(\bx_0)>0}
\lambda_i(\bx_0)
\label{eq:endmatter-covariance-growth}
\end{equation}
for almost every initial condition $\bx_0$ under the conditions stated below.
Substituting Eq.~\eqref{eq:endmatter-covariance-growth} into Eq.~\eqref{eq:finiteV-gaussian-information-loss} and interchanging the long-time limit with the stationary average gives Eq.~\eqref{eq:covariance-lyapunov-rate}.

We fix an initial concentration $\bm{X}_0^{(V)}=\bx_0$ and let $\bm{\phi}_t(\bx_0)$ be the corresponding deterministic trajectory.
Define the scaled fluctuation by
$\bm{\xi}_t\equiv\sqrt{V}\,[\bm{X}_t^{(V)}-\bm{\phi}_t(\bx_0)]$.
To leading order in the system-size expansion, and for times over which the fluctuations remain sufficiently small that the drift can be linearized about the deterministic trajectory, $\bm{\xi}_t$ obeys a linear Langevin equation. Its covariance $\bm{\Sigma}_t(\bx_0)$ therefore satisfies
\begin{equation}
\dot{\bm{\Sigma}}_t
=
\bm{A}_t\bm{\Sigma}_t
+
\bm{\Sigma}_t\bm{A}_t^{\mathsf T}
+
\bm{D}_t,
\qquad
\bm{\Sigma}_0=0,
\label{eq:linear-covariance}
\end{equation}
where
\begin{equation}
\bm{A}_t
=
D\bF\!\left(\bm{\phi}_t(\bx_0)\right),
\qquad
\bm{D}_t
=
\sum_{r=1}^{R}
a_r\!\left(\bm{\phi}_t(\bx_0)\right)
\bnu_r\bnu_r^{\mathsf T}.
\label{eq:endmatter-linear-coefficients}
\end{equation}
Here, $D\bF(\by)$ denotes the Jacobian of $\bF$ with components $[D\bF(\by)]_{ij}=\partial F_i(\by)/\partial y_j$, and $\bm{D}_t$ is the diffusion matrix evaluated along the deterministic trajectory.

Let $\bm{\Phi}(t,s)$ denote the propagator of the linearized deterministic flow along
this trajectory:
\begin{equation}
\partial_t\bm{\Phi}(t,s)
=
\bm{A}_t\bm{\Phi}(t,s),
\qquad
\bm{\Phi}(s,s)=\bm{1}.
\label{eq:endmatter-propagator}
\end{equation}
The solution of Eq.~\eqref{eq:linear-covariance} is
\begin{equation}
\bm{\Sigma}_t(\bx_0)
=
\int_0^t
\bm{\Phi}(t,s)
\bm{D}_s
\bm{\Phi}(t,s)^{\mathsf T}
\,ds.
\label{eq:endmatter-gramian}
\end{equation}

Using the composition rule
$\bm{\Phi}(t,s)=\bm{\Phi}(t,0)\bm{\Phi}(s,0)^{-1}$,
Eq.~\eqref{eq:endmatter-gramian} can be factorized on the full tangent space as
\begin{equation}
\bm{\Sigma}_t
=
\bm{\Phi}(t,0)
\bm{R}_t
\bm{\Phi}(t,0)^{\mathsf T},
\label{eq:endmatter-full-factorization}
\end{equation}
where
\begin{equation}
\bm{R}_t
=
\int_0^t
[\bm{\Phi}(s,0)]^{-1}
\bm{D}_s
[\bm{\Phi}(s,0)]^{-\mathsf T}
\,ds.
\label{eq:endmatter-full-R}
\end{equation}
Taking the determinant of Eq.~\eqref{eq:endmatter-full-factorization} gives the exact identity
\begin{equation}
\ln\det\bm{\Sigma}_t
=
2\ln\left|
\det\bm{\Phi}(t,0)
\right|
+
\ln\det\bm{R}_t.
\label{eq:endmatter-full-determinant}
\end{equation}

We now introduce the Lyapunov decomposition at $t=0$.
Let $E_0$ be the tangent space at $\bx_0$, and choose covariant Lyapunov vectors $\bm{v}_i(0)$ adapted to the splitting
\begin{equation}
E_0
=
E_0^u
\oplus
E_0^0
\oplus
E_0^s,
\label{eq:endmatter-lyapunov-splitting}
\end{equation}
where the three subspaces correspond to positive, zero, and negative Lyapunov exponents, respectively.
Their defining covariant property under the linearized flow means that normalized vectors $\bm{v}_i(t)$ and positive stretching factors $\gamma_i(t)$ can be chosen so that~\cite{Gaspard1998}
\begin{equation}
\bm{\Phi}(t,0)\bm{v}_i(0)
=
\gamma_i(t)\bm{v}_i(t),
\label{eq:endmatter-covariant-lyapunov-growth}
\end{equation}
where Oseledets' theorem gives
\[
\lim_{t\to\infty}
\frac{1}{t}\ln\gamma_i(t)
=
\lambda_i(\bx_0)
\]
for almost every initial concentration $\bx_0$~\cite{Oseledets}.
Provided that the angles between the Lyapunov subspaces do not close exponentially, the determinant of the propagator satisfies
\begin{equation}
\lim_{t\to\infty}
\frac{1}{t}
\ln\left|
\det\bm{\Phi}(t,0)
\right|
=
\sum_i\lambda_i(\bx_0).
\label{eq:endmatter-full-volume-growth}
\end{equation}

We next evaluate the second term in Eq.~\eqref{eq:endmatter-full-determinant} in the covariant Lyapunov basis of $E_0$.
Let $d_i(s)$ denote the diffusion coefficient projected onto the $i$th Lyapunov coordinate using the dual covariant basis.
Equation~\eqref{eq:endmatter-full-R} shows that the corresponding diagonal component $r_i(t)$ of $\bm{R}_t$ is
\begin{equation}
r_i(t)
=
\int_0^t
\gamma_i(s)^{-2}d_i(s)
\,ds.
\label{eq:endmatter-pulled-back-direction}
\end{equation}
Since $\gamma_i(s)=\exp[\lambda_i s+o(s)]$, its exponential rate follows directly from this integral.
If $d_i(s)$ is bounded above and bounded away from zero, then
\begin{equation}
\lim_{t\to\infty}
\frac{1}{2t}\ln r_i(t)
=
\begin{cases}
0,
& \lambda_i\geq 0,
\\
-\lambda_i,
& \lambda_i<0,
\end{cases}
=
\max\{-\lambda_i,0\}.
\label{eq:endmatter-pulled-back-growth}
\end{equation}
Thus, $r_i(t)$ remains bounded for an unstable direction, grows subexponentially for a neutral direction, and grows as $e^{-2\lambda_i t}$ for a stable direction.
Assuming that the diffusion generates a nonzero spread in every Lyapunov direction and that correlations between nonorthogonal covariant directions do not change the leading determinant rate, we obtain
\begin{equation}
\lim_{t\to\infty}
\frac{1}{2t}\ln\det\bm{R}_t
=
-\sum_{\lambda_i(\bx_0)<0}
\lambda_i(\bx_0).
\label{eq:endmatter-R-growth}
\end{equation}

Equations~\eqref{eq:endmatter-full-determinant},
\eqref{eq:endmatter-full-volume-growth}, and
\eqref{eq:endmatter-R-growth} therefore give
\begin{align}
\lim_{t\to\infty}
\frac{1}{2t}\ln\det\bm{\Sigma}_t(\bx_0)
&=
\sum_i\lambda_i(\bx_0)
-
\sum_{\lambda_i(\bx_0)<0}\lambda_i(\bx_0)
\notag\\
&=
\sum_{\lambda_i(\bx_0)>0}
\lambda_i(\bx_0),
\end{align}
which is Eq.~\eqref{eq:endmatter-covariance-growth}.
In particular, a stable direction contributes $\lambda_i<0$ through the deterministic factor in Eq.~\eqref{eq:endmatter-full-determinant}, whereas the continuous injection of reaction noise contributes $-\lambda_i$ through $\bm{R}_t$; the two rates cancel exactly.
Without continuous noise, a nonsingular initial covariance would instead be propagated only by $\bm{\Phi}(t,0)$, and the stable direction would retain its negative Lyapunov exponent.

\newpage
\begin{figure}[H]
  \centering
  \includegraphics[
    width=\linewidth
  ]{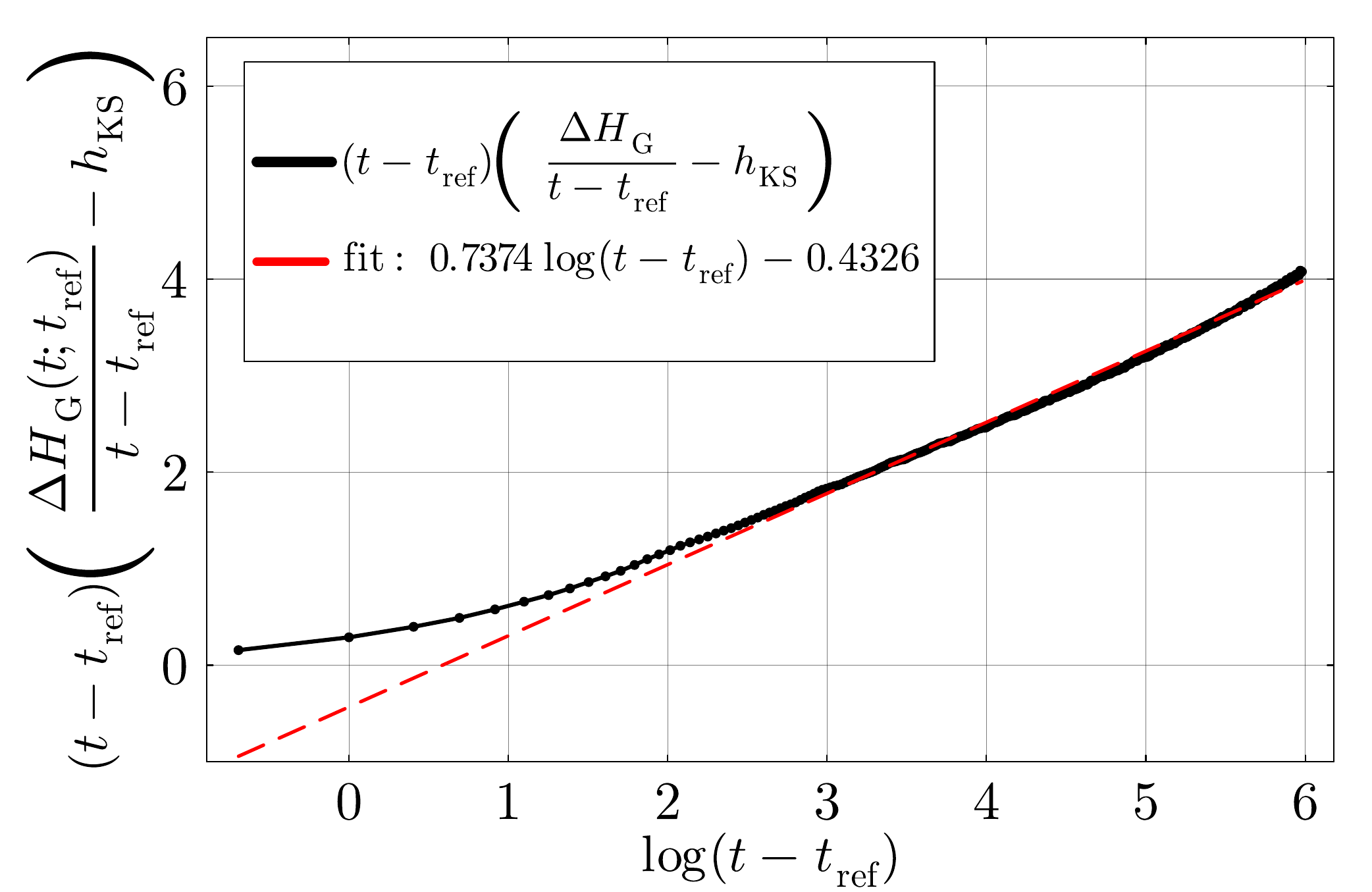}
  \caption{
      Finite-time correction to the rate of Gaussian information loss for the modified CS14 model.
      The black curve shows
      $(t-t_{\mathrm{ref}})
      [\Delta H_{\mathrm G}^{(V)}(t;t_{\mathrm{ref}})/(t-t_{\mathrm{ref}})-h_{\mathrm{KS}}]$
      as a function of $\ln(t-t_{\mathrm{ref}})$.
      The red dashed line shows a linear fit.
      The approximately linear dependence is consistent with a correction proportional to
      $\ln(t-t_{\mathrm{ref}})/(t-t_{\mathrm{ref}})$.
  }
  \label{fig:endmatter-logarithmic-correction}
\end{figure}
\section{Finite-time correction to \NoCaseChange{Eq.~\eqref{eq:covariance-lyapunov-rate}}}
Figure~\ref{fig:endmatter-logarithmic-correction} shows the finite-time approach of the Gaussian entropy-growth rate to the KS entropy.
The approximately linear dependence of the transformed residual on $\ln(t-t_{\mathrm{ref}})$ indicates a correction proportional to $\ln(t-t_{\mathrm{ref}})/(t-t_{\mathrm{ref}})$.
To determine its origin, we examine the finite-time behavior implied by Eq.~\eqref{eq:endmatter-pulled-back-direction}.
The modified CS14 model has one zero Lyapunov exponent.
For its neutral direction, we assume that the corresponding diagonal component of $\bm{R}_t$ exhibits power-law growth, $r_0(t)\sim t^{\alpha_0}$, with $\alpha_0>0$.
These assumptions give the following asymptotic forms:
\begin{equation}
\frac{1}{2t}\ln r_i(t)
=
\begin{cases}
\mathcal{O}(t^{-1}),
& \lambda_i>0,
\\[1ex]
\displaystyle
\frac{\alpha_0\ln t}{2t}
+
\mathcal{O}(t^{-1}),
& \lambda_i=0,
\\[2ex]
-\lambda_i+o(1),
& \lambda_i<0.
\end{cases}
\label{eq:endmatter-pulled-back-growth}
\end{equation}
Among these directions, only the neutral one produces a logarithmic correction to the Gaussian entropy-growth rate.
For fixed $t_{\mathrm{ref}}>0$, the second line of Eq.~\eqref{eq:endmatter-pulled-back-growth} gives a contribution
$\alpha_0\ln(t-t_{\mathrm{ref}})/[2(t-t_{\mathrm{ref}})]$
up to terms of order $(t-t_{\mathrm{ref}})^{-1}$.
If all remaining corrections to the Gaussian entropy-growth rate are of order $(t-t_{\mathrm{ref}})^{-1}$, then
\begin{align}
&(t-t_{\mathrm{ref}})
\left[
\frac{
\Delta H_{\mathrm G}^{(V)}(t;t_{\mathrm{ref}})
}{
t-t_{\mathrm{ref}}
}
-
h_{\mathrm{KS}}
\right]
\notag\\
&\qquad =
\frac{\alpha_0}{2}\ln(t-t_{\mathrm{ref}})
+
\mathcal{O}(1).
\label{eq:endmatter-neutral-finite-time-correction}
\end{align}
The observed logarithmic correction is therefore consistent with the assumed power-law growth of $r_0(t)$.

\end{document}